\documentclass[11pt]{article}
\usepackage{amsmath}
\usepackage{amssymb}
\usepackage{fontspec}
\usepackage{wasysym}
\usepackage[pdfusetitle,
 bookmarks=true,bookmarksnumbered=false,bookmarksopen=false,
 breaklinks=false,pdfborder={0 0 0},pdfborderstyle={},backref=false,colorlinks=false]
 {hyperref}

\makeatletter

\providecommand{\tabularnewline}{\\}

\usepackage{amsfonts}\usepackage{graphicx}
\usepackage{graphics}
\usepackage{eepic}
\usepackage{epsfig}

\@addtoreset{equation}{section}

\makeatother

\begin{document}
\title{Scalar vacuum densities on Beltrami pseudosphere}
\author{T. A. Petrosyan\thanks{E-mail: tigran.petrosyan@ysu.am} \vspace{0.3cm}
 \\
 %EndAName
\textit{Institute of Physics, Yerevan State University,}\\
 \textit{1 Alex Manoogian Street, 0025 Yerevan, Armenia} \vspace{0.3cm}
}
\maketitle
\begin{abstract}
We investigate the combined effects of spatial curvature and topology
on the properties of the vacuum state for a charged scalar field localized
on the (2+1)-dimensional Beltrami pseudosphere, assuming that the
field obeys quasiperiodicity condition with constant phase. As important
local characteristics of the vacuum state the vacuum expectation values
(VEVs) of the field squared and energy-momentum tensor are evaluated.
The contributions in the VEVs coming from geometry with an uncompactified
azimuthal coordinate are divergent, whereas the compact counterparts
are finite and are analysed both numerically and asymptotically. For
small values of proper radius of the compactified dimension, the leading
terms of topological contributions are independent of the field mass
and curvature coupling parameter, increasing by a power-law. In the
opposite limit, the VEVs decay following a power-law in the general
case. In the special case of a conformally coupled massless field
the behavior is different. Unlike the VEV of field squared and vacuum
energy density, the radial and azimuthal stresses are increasing by
absolute value. As a consequence, the effects of nontrivial topology
are strong for the stresses in this case at small values of radial
coordinate.
\end{abstract}

\section{Introduction}

\label{sec:Int}

The investigation of field-theoretical effects in (2+1)-dimensional
curved spacetimes is motivated by several reasons \cite{Birr82,Grib94,Park09}.
In backgrounds of this dimensionality the underlying laws of a system
sometimes possess a certain symmetry, but the physically realized
state of the system does not exhibit that symmetry. Particularly,
there can occur phenomena where the laws of physics are not invariant
under spatial inversion. Moreover, the strong correlations and topological
effects in quantum many-body systems can cause fractionalization of
the fundamental quantum numbers for the observable excitations. In
addition, mechanisms can be provided for the generation of a topological
mass for gauge bosons which is not accompanied with gauge symmetry
breaking. There are compelling motivations in the context of cosmological
research as well, such as the effective field theories describing
the physics of compact subspace in Kaluza-Klein and braneworld models
with extra dimensions. The Kaluza-Klein theory employs dimensional
reduction by introducing an additional compactified spatial dimension,
which allows for the unification of electromagnetism and gravity.
The Kaluza-Klein dimensional reduction technic is employed for 2D
elastic plates in \cite{Adda23}. The continuous development of planar
condensed matter systems, such as carbon nanostructures, has increased
significantly the interest towards 2D models. The long-wavelength
degrees of freedom in those systems are described by 2D Dirac theory
\cite{Gusy07,Cast09}.

In many two-dimensional quantum field theory models, including those
relevant to condensed matter physics, the physical degrees of freedom
are confined within finite regions by applying specific periodicity
and boundary conditions. When topological defects are present, additional
constraints on the fields are also introduced. Analyzing how boundary
conditions influence quantum fields in (2+1) dimensions helps develop
more realistic and practically useful models. Moreover, studying edge-related
effects in (2+1)-dimensional spacetime offers valuable insights into
the anti-de Sitter/conformal field theory (AdS/CFT) correspondence,
where a (2+1)-dimensional quantum theory resides on the boundary of
a (3+1)-dimensional AdS space. This perspective deepens our understanding
of holography and its connection to string theory and quantum gravity.
The boundary conditions alter the spectrum of quantum fluctuations,
leading to boundary-dependent modifications in the expectation values
of physical quantities. This shift is known as the Casimir effect,
extensively studied for its fundamental importance and applications
in nanoscale physics. Similar phenomena occur in systems with compact
spatial dimensions, where the compactification of fields gives rise
to topological Casimir effects, observed across a wide range of boundary
and geometric configurations. The investigation of the nontrivial
topology in (2+1)-dimensional models is currently a subject of active
research. In particular, the 2D materials having negative constant
curvature have various interesting condensed matter realizations.
Recently, such materials have been used as simplified models for 2D
black holes and wormholes \cite{Iori12,Iori14,Gall21,Gall22,Alen21}.
New kinds of structures can be obtained by combining different geometries
with rotational symmetry. For example, capped graphene nanotubes combine
spherical and cylindrical geometries in their structure. The vacuum
expectation values (VEVs) of the field squared and of the energy-momentum
tensor for a quantum scalar field on that background are investigated
in \cite{Beze16}. The generation of persistent currents on 2D curved
tubes is a well-known effect \cite{Dunn99,Imry08,Fomi18,Saha24a,Bles09,Bluh09}.

Rotationally symmetric (2+1)-dimensional spacetimes give a manageable
setting to explore how mass, rotation, and curvature interact in general
relativity \cite{Saha24,Saha25a,Iori15}. The general (2+1)-dimensional
rotationally symmetric spacetime can be described in coordinates $(t,w,\phi)$
by the line element $ds^{2}=dt^{2}-dw^{2}-p^{2}(w)d\phi^{2}$, where
$p(w)$ is some function of radial coordinate $w$. The simplest special
case of this geometry corresponds to a cylinder with a constant radius
(in this case $p(w)$ is equal to the cylinder radius). If the function
$p(w)$ is linear, $p(w)=\alpha w$, then one has either a plane ($\alpha=1$)
or a cone with an opening angle $2\pi\alpha$ ($\alpha<1$). The fermionic
condensate and currents for a 2D fermionic field localized on a conical
spacetime with two parallel circular boundaries are discussed in \cite{Saha21,Bell20}
(see also \cite{Saha25} for finite temperature fermionic currents
in case of one boundary). The gravitational response of non-interacting 2D fermions in
an external magnetic field is evaluated in \cite{Aban14}, where the
contributions of conic singularity in the expectation values of the charge
density and the energy--momentum tensor are discussed. The constant positive curvature space is
another example of the general geometry corresponding to a sphere
of radius $a$. In the latter case one has $p(w)=L\sin(w/a)$, with
$L$ being a constant having the dimension of length. The corresponding
condensed matter realizations include fullerenes and topological insulators
with spherical surfaces. The VEVs of the field squared and of the
energy-momentum tensor for a complex scalar field on the background
of de Sitter spacetime and Milne universe in the presence of a spherical
surface are studied in \cite{Saha21a,Saha22,Saha20}. As another special
case the constant negative curvature spaces can be mentioned. In this
case three different subcases can be realized: the elliptic pseudosphere
($p(w)=L\sinh(w/a)$), the hyperbolic pseudosphere ($p(w)=L\cosh(w/a)$)
and the Beltrami pseudosphere ($p(w)=Le^{w/a}$). The solution of
Einstein-Maxwel equations with negative cosmological constant in (2+1)-dimensions
corresponds to the lower-dimensional analog of a black hole, known
as BTZ black hole. The latter is described by a rotationally symmetric
spacetime with constant negative curvature. The corresponding black
hole entropy and temperature for the cases of vanishing and nonzero
charges are evaluated in \cite{Bana92} (see also \cite{Brow86,Carl95,Carl98,Stro98}).
It is shown that for negative values of the mass there is a naked
conical singularity. However, it disappears in the special case of
vanishing angular momentum, provided that the mass is precisely $-1$.
In this bound state the line element coincides with that of the (2+1)-dimensional
AdS spacetime. The investigation of BTZ black hole provides insight
especially in quantum properties of black holes. The renormalized
energy-momentum tensor for strongly coupled quantum conformal fields
in the rotating BTZ black hole is presented in \cite{Empa20}. In
the present paper we consider the (2+1)-dimensional Beltrami pseudosphere
as a background geometry, and we are interested in the effects induced
by the spatial topology and curvature on the vacuum characteristics
of a massive complex scalar field. The focus of the present paper is
the determination of VEVs of the field squared and the energy--momentum tensor,
whereas the vacuum current density is evaluated in \cite{Saha24}.

The paper is organized as follows. In the next section we describe
the background geometry and present the expressions for the Hadamard
function. By using the Hadamard function, the VEV of the field squared
is evaluated in Section \ref{sec:fldsq}. The corresponding expressions
for the diagonal components of the VEV of energy-momentum tensor,
as well as the asymptotic and numerical analysis of the corresponding
topological contributions is presented in Section \ref{sec:emt}.
The main results of the paper are summarized in Section \ref{sec:conc}.

\section{Problem geometry and Hadamard function}

\label{sec:Had}

\begin{figure}[tbph]
\begin{centering}
\epsfig{figure=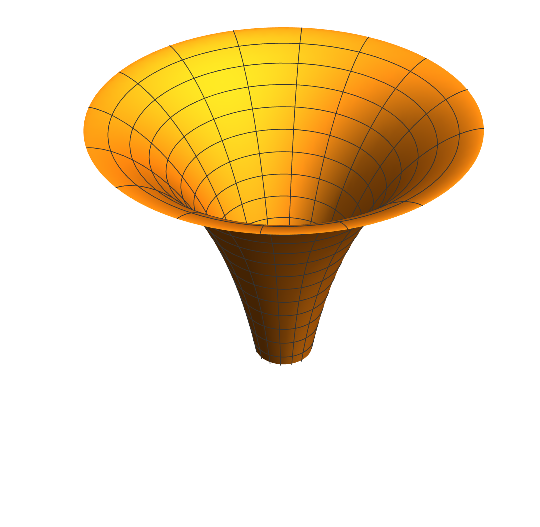,width=8cm,height=7cm}
\par\end{centering}
\caption{The Beltrami pseudosphere embedded in a 3-dimensional Euclidean space.}
\label{fig1}
\end{figure}

The (2+1)-dimensional Beltrami pseudosphere is described by the spacetime
coordinates $(t,r,\phi)$ with the line element \cite{Saha24}
\begin{equation}
ds^{2}=dt^{2}-\frac{a^{2}}{r^{2}}\left(dr^{2}+L^{2}d\phi^{2}\right),\label{ds2}
\end{equation}
where $a$ is the curvature radius and $L$ is a constant with dimension
of length. The parameter $L$ is physically relevant for deformed graphene
systems and is related to the graphene lattice spacing. The radial coordinate is related to the coordinate $w$
mentioned in the Section \ref{sec:Int} as $r=ae^{-w/a}$, $0\leq r<\infty$.
For the part of the manifold that can be embedded in a 3-dimensional
Euclidean space one has $r\geq L$. Note that the maximal circle at $r=L$, known as the Hilbert horizon of the
Beltrami pseudosphere, corresponds to the event horizon of the BTZ black hole in
the case of zero mass and zero angular momentum \cite{Iori14}. Thus, the
minimal radial coordinate $L$ can be associated with a "throat" size in a
wormhole analogy. For a given $r$ the
ratio $aL/r$ corresponds to the proper radius of the compactified
dimension. Varying $r$ at fixed $L$ changes the
proper circumference of the compact dimension, making $L/r$ the natural
expansion parameter of the problem. The nonzero components of the Ricci tensor and the Ricci
scalar have the form 
\begin{equation}
R_{1}^{1}=R_{2}^{2}=-\frac{1}{a^{2}},\;R=-\frac{2}{a^{2}}.\label{RBirr}
\end{equation}
The Beltrami pseudosphere depicted in Fig. \ref{fig1} represents
a constant negative curvature space with the Gaussian curvature $K=R/2$.

We consider a massive scalar field $\varphi(x)\equiv\varphi(t,r,\phi)$
on the background of the Beltrami pseudosphere. The field obeys the
equation 
\begin{equation}
\left(\nabla^{p}\nabla_{p}+m^{2}+\xi R\right)\varphi(x)=0,\label{KGeqn}
\end{equation}
where $m$ is the field mass, $\xi$ is the curvature coupling parameter
and $\nabla_{p}$ is the covariant derivative operator. The field
obeys also the quasiperiodicity condition
\begin{equation}
\varphi(t,r,\phi+2\pi)=e^{i\alpha_{p}}\varphi(t,r,\phi),\label{pc}
\end{equation}
with a constant phase $\alpha _{p}$ having values in the range $0\leq \alpha
_{p}<2\pi $. The latter can be interpreted
physically as the existence of a magnetic flux in the embedding space,
threading the Beltrami pseudosphere (specifically, the magnetic flux multiplied by $2\pi $, in units of the flux
quantum). The corresponding two-point Hadamard
function has the form \cite{Saha24} 
\begin{equation}
G(x,x^{\prime})=\frac{a^{-2}}{2\pi}\sum_{l=-\infty}^{\infty}e^{-il\alpha_{p}}\int_{0}^{\infty}d\nu\,\frac{\nu}{\omega}\tanh\left(\pi\nu\right)\cos(\omega\Delta t)\,P_{i\nu-1/2}\left(\frac{r^{2}+r^{\prime2}+L^{2}\left(\Delta\phi+2\pi l\right)^{2}}{2rr^{\prime}}\right),\label{Gcc}
\end{equation}
where $P_{\mu}(z)$ is the Legendre function of the first kind, $\Delta\phi=\phi-\phi^{\prime}$,
$\Delta t=t-t^{\prime}$, $\omega=\sqrt{\nu^{2}+\nu_{m}^{2}}/a$ and
$\nu_{m}^{2}=m^{2}a^{2}+1/4-2\xi$. The term with $l=0$ in (\ref{Gcc})
corresponds to the Hadamard function in the geometry with an uncompactified
coordinate $\phi$. The latter part is given as
\begin{equation}
G_{0}(x,x^{\prime})=\frac{1}{2\pi a}\int_{0}^{\infty}d\nu\,\frac{\nu}{\sqrt{\nu^{2}+\nu_{m}^{2}}}\tanh(\nu\pi)\cos(\omega\Delta t)P_{i\nu-1/2}\left(\frac{r^{2}+r^{\prime2}+(L\Delta\phi)^{2}}{2rr^{\prime}}\right).\label{G10cc1}
\end{equation}
As an alternative representation for the Hadamard function we use
the expression
\begin{equation}
G(x,x^{\prime})=\frac{1}{\pi^{2}a}\sum_{l=-\infty}^{\infty}e^{il\alpha_{p}}\int_{\nu_{m}}^{\infty}d\nu\,\frac{\nu\cosh\left(\sqrt{\nu^{2}-\nu_{m}^{2}}\Delta t/a\right)}{\sqrt{\nu^{2}-\nu_{m}^{2}}}Q_{\nu-1/2}\left(\frac{r^{2}+r^{\prime2}+L^{2}\left(\Delta\phi-2\pi l\right)^{2}}{2rr^{\prime}}\right),\label{Gccn4}
\end{equation}
where again the Hadamard function in the geometry with uncompactified
coordinate $\phi$, $G_{0}(x,x^{\prime})$, is given by the term with
$l=0$. Here $Q_{\mu}(z)$ is the Legendre function of the second
kind \cite{Abra72}. The Hadamard function for a general rotationally symmetric geometry
can be obtained by the mode summation method by applying the generalized
Abel-Plana formula \cite{Saha24,Saharev2}.

\section{VEV of the field squared}

\label{sec:fldsq}

Having the expression for the Hadamard function we can evaluate the
characteristics of the scalar vacuum. In this section we will investigate
the VEV of the field squared which can be found by the formula
\begin{equation}
\left\langle \varphi^{2}\right\rangle =\frac{1}{2}\underset{x^{\prime}\rightarrow x}{\lim}G\left(x,x^{\prime}\right).\label{FldSq0}
\end{equation}
Substituting here (\ref{Gcc}) we get
\begin{equation}
\left\langle \varphi^{2}\right\rangle =\frac{1}{4\pi a}\int_{0}^{\infty}d\nu\,\frac{\nu\tanh\left(\pi\nu\right)}{\sqrt{\nu^{2}+\nu_{m}^{2}}}+\frac{1}{2\pi a}\sum_{l=1}^{\infty}\cos\left(l\alpha_{p}\right)\int_{0}^{\infty}d\nu\,\frac{\nu}{\sqrt{\nu^{2}+\nu_{m}^{2}}}\tanh\left(\pi\nu\right)P_{i\nu-1/2}\left(x\right),\label{FldSq}
\end{equation}
where $x$ is defined as
\begin{equation}
x\equiv x_{l}\left(r\right)=1+2\left(\frac{\pi lL}{r}\right)^{2},\:l\neq0.\label{xr}
\end{equation}
The first term in (\ref{FldSq}), which is divergent, corresponds
to the geometry with an uncompactified azimuthal coordinate and it
does not depend on the compactification length and radial coordinate.
The dependence on $L$ and $r$ is contained in the last term of (\ref{FldSq})
and it appears in the form of the ratio $L/r$. The latter presents
the proper length of the compact dimension measured in units of $a$.
Hence the covariant d'Alembertian of the VEV has the form
\begin{equation}
\nabla_{p}\nabla^{p}\left\langle \varphi^{2}\right\rangle =\frac{1}{2\pi a^{3}}\sum_{l=1}^{\infty}\cos\left(l\alpha_{p}\right)\int_{0}^{\infty}d\nu\,\frac{\nu}{\sqrt{\nu^{2}+\nu_{m}^{2}}}\tanh\left(\pi\nu\right)\hat{F}_{\Square}P_{i\nu-1/2}\left(x\right),\label{covdal}
\end{equation}
with the operator $\hat{F}_{\Square}$ defined as
\begin{equation}
\hat{F}_{\Square}=-2\left(x-1\right)\left[2\left(x-1\right)\partial_{x}^{2}+3\partial_{x}\right].\label{Fcovdal}
\end{equation}
An alternative representation suitable for numerical evaluations can
be obtained by using (\ref{Gccn4}). The latter gives
\begin{equation}
\left\langle \varphi^{2}\right\rangle =\frac{1}{2\pi^{2}a}\int_{\nu_{m}}^{\infty}d\nu\,\frac{\nu}{\sqrt{\nu^{2}-\nu_{m}^{2}}}\lim_{x\rightarrow1}Q_{\nu-1/2}\left(x\right)+\left\langle \varphi^{2}\right\rangle _{\mathrm{c}},\label{FldSq1}
\end{equation}
where the first term is again divergent and needs a renormalization.
The second term on the right hand side of (\ref{FldSq1}) corresponds
to the compact counterpart in the VEV of field squared and has the
form
\begin{equation}
\left\langle \varphi^{2}\right\rangle _{\mathrm{c}}=\frac{1}{\pi^{2}a}\sum_{l=1}^{\infty}\cos\left(l\alpha_{p}\right)\int_{\nu_{m}}^{\infty}d\nu\,\frac{\nu}{\sqrt{\nu^{2}-\nu_{m}^{2}}}Q_{\nu-1/2}\left(x\right).\label{FldSq2}
\end{equation}
In this form for the corresponding covariant d'Alembertian we obtain
\begin{equation}
\nabla_{p}\nabla^{p}\left\langle \varphi^{2}\right\rangle =\frac{1}{\pi^{2}a^{3}}\sum_{l=1}^{\infty}\cos\left(l\alpha_{p}\right)\int_{\nu_{m}}^{\infty}d\nu\,\frac{\nu}{\sqrt{\nu^{2}-\nu_{m}^{2}}}\hat{F}_{\Square}Q_{\nu-1/2}\left(x\right),\label{covdal-1}
\end{equation}
with the same operator $\hat{F}_{\Square}$ defined as (\ref{Fcovdal}).
In this paper our main interest are the topological contributions
to the VEVs of field squared and energy-momentum tensor.

As another characteristic of the vacuum state we can consider the
VEV of the current density $\langle0|j_{k}(x)|0\rangle\equiv\langle j_{k}(x)\rangle$,
where 
\begin{equation}
j_{k}(x)=ie[\varphi^{\dagger}(x)\partial_{k}\varphi(x)-(\partial_{k}\varphi(x))^{\dagger}\varphi(x)]\label{jl}
\end{equation}
is the corresponding operator and $e$ is the charge of the field
quantum. This VEV can be obtained from the Hadamard function by the
formula
\begin{equation}
\langle j_{k}(x)\rangle=\frac{i}{2}e\lim_{x^{\prime}\rightarrow x}(\partial_{k}-\partial_{k}^{\prime})G(x,x^{\prime}).\label{jl1}
\end{equation}
According to \cite{Saha24} the only nonzero component is the VEV
of the current density in the compact dimension $\langle j_{2}\rangle$.
The corresponding physical component $\langle j^{\phi}\rangle=-r\langle j_{2}\rangle/La$ (the metric signature $(+,-,-)$ is used everywhere
throughout the paper) is given as
\begin{equation}
\langle j^{\phi}\rangle=-\frac{4eL}{\pi ra^{2}}\sum_{l=1}^{\infty}l\sin\left(l\alpha_{p}\right)\int_{\nu_{m}}^{\infty}d\nu\,\nu\frac{Q_{\nu-1/2}^{\prime}\left(1+2\left(\pi lL/r\right)^{2}\right)}{\sqrt{\nu^{2}-\nu_{m}^{2}}},\label{jphicc3}
\end{equation}
where the prime means the derivative with respect to the argument
of the function. As seen from (\ref{FldSq2}) and (\ref{jphicc3}),
the VEV of the field squared is an even function of $\alpha_{p}$,
whereas the VEV of the current density is an odd function.

Let us consider the behavior of $\left\langle \varphi^{2}\right\rangle _{\mathrm{c}}$
in the asymptotic regions of the ratio $r/L$. For $r/L\ll1$ the
argument of the Legendre function in (\ref{FldSq2}) is large and
we use the formula \cite{Olve10}
\begin{equation}
Q_{\nu-1/2}\left(x\right)\approx\frac{\sqrt{\pi}\Gamma\left(\nu+1/2\right)}{\Gamma\left(\nu+1\right)\left(2x\right)^{\nu+1/2}},\,x\gg1.\label{Ql}
\end{equation}
In the leading order this gives
\begin{equation}
\left\langle \varphi^{2}\right\rangle _{\mathrm{c}}\approx\frac{\Gamma\left(\nu_{m}+1/2\right)}{2\pi a\Gamma\left(\nu_{m}\right)\sqrt{\nu_{m}\ln\left(2\pi L/r\right)}}\left(\frac{r}{2\pi L}\right)^{2\nu_{m}+1}\sum_{l=1}^{\infty}\frac{\cos\left(l\alpha_{p}\right)}{l^{2\nu_{m}+1}},\label{smrfs}
\end{equation}
and $\left\langle \varphi ^{2}\right\rangle _{\mathrm{c}}$ tends to zero
like $\left( r/L\right) ^{2\nu _{m}+1}$. This expression is obtained by
taking into account that in the integral in (\ref{FldSq2}) the main
contribution comes from the lower limit of integration, and the logarithmic
factor appears as a result of integration over $\nu $. Analogically, the
corresponding leading term in case of a conformally coupled massless field ($%
\nu _{m}=0$) is given~as%
\begin{equation}
\left\langle \varphi^{2}\right\rangle _{\mathrm{c}}\approx\frac{1}{4\pi^{2}a}\frac{1}{\ln\left(2\pi L/r\right)}\frac{r}{L}\sum_{l=1}^{\infty}\frac{1}{l}\cos\left(l\alpha_{p}\right).\label{smrfsc}
\end{equation}
Note that in the limit under consideration $L/r\gg1$ and this corresponds
to large values of the proper length of compact dimension compared
to the curvature radius $a$.

In the opposite limit $r/L\gg1$ the dominant contribution to the
integral in (\ref{FldSq2}) comes from large $\nu$ and we use the
uniform asymptotic approximation \cite{Olve10}
\begin{equation}
Q_{\nu-1/2}(\cosh u)\approx\left(\frac{u}{\sinh u}\right)^{1/2}K_{0}\left(\nu u\right),\;\nu\gg1.\label{AsLeg}
\end{equation}
After that we perform integration by $\nu$ (see \cite{Prud2}), and for the leading order
term we obtain

\begin{equation}
\left\langle \varphi^{2}\right\rangle _{\mathrm{c}}\approx\frac{r/L}{4\pi^{2}a}\sum_{l=1}^{\infty}\frac{1}{l}\cos\left(l\alpha_{p}\right).\label{lgrfldsq}
\end{equation}
This result coincides with the corresponding VEV for the cylindrical
tube with a constant radius $aL/r$. The dominant contribution comes
from the modes with high momenta along the compact dimension and the
influence of the curvature on those modes is weak.

\begin{figure}[tbph]
\begin{centering}
\begin{tabular}{cc}
\epsfig{figure=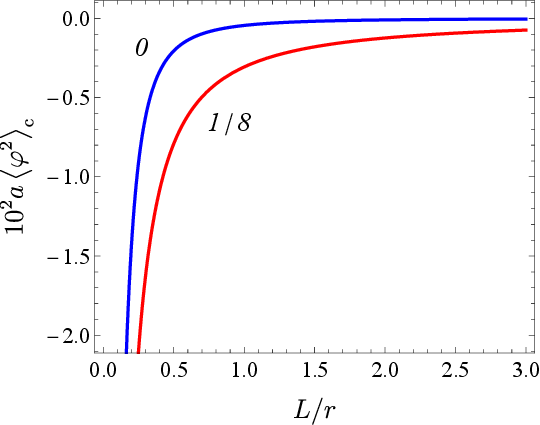,width=7.5cm,height=6cm} & \quad{}\epsfig{figure=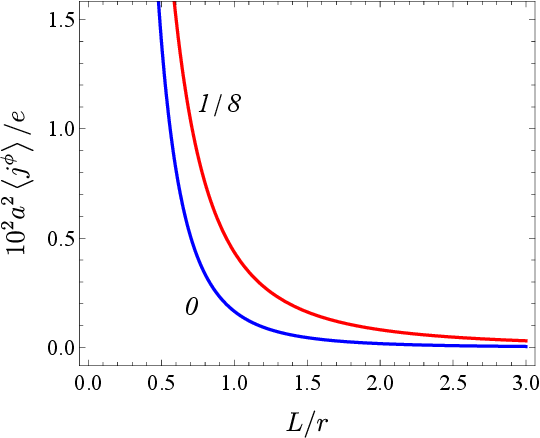,width=7.5cm,height=6cm}\tabularnewline
\end{tabular}
\par\end{centering}
\caption{The VEV of field squared (left panel) and the azimuthal current density
(right panel) as a function of the $L/r$ for conformally ($\xi=1/8$)
and minimally ($\xi=0$) coupled massless fields and for fixed $\alpha_{p}=\pi/2$. The numbers near the curves are the values of $\protect\xi .$}
\label{fig2}
\end{figure}

\begin{figure}[tbph]
\begin{centering}
\begin{tabular}{cc}
\epsfig{figure=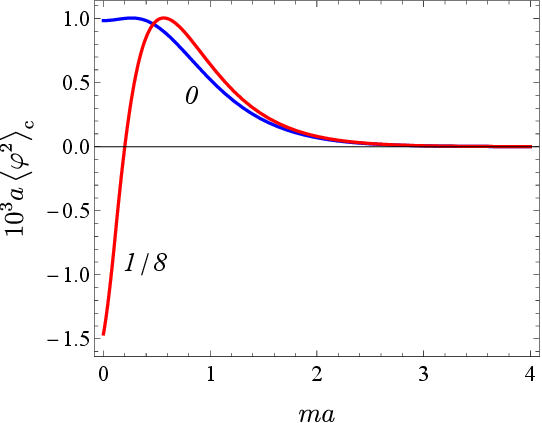,width=7.5cm,height=6cm} & \quad{}\epsfig{figure=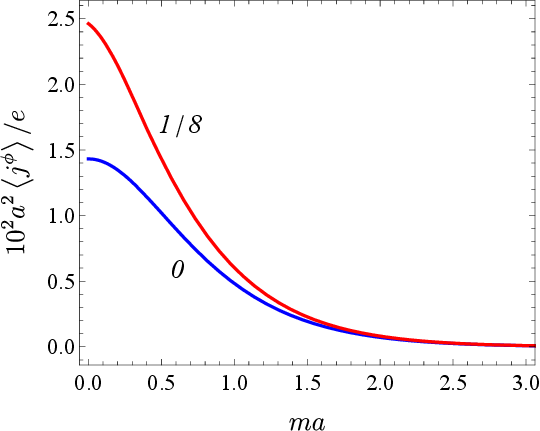,width=7.5cm,height=6cm}\tabularnewline
\end{tabular}
\par\end{centering}
\caption{The VEV of field squared (left panel) and the azimuthal current density
(right panel) as a function of the mass for conformally ($\xi=1/8$)
and minimally ($\xi=0$) coupled fields with $L/r=0.5$ and $\alpha_{p}=2\pi/5$. The numbers near the curves are the values of $\protect\xi .$}
\label{fig3}
\end{figure}

\begin{figure}[tbph]
\begin{centering}
\begin{tabular}{cc}
\epsfig{figure=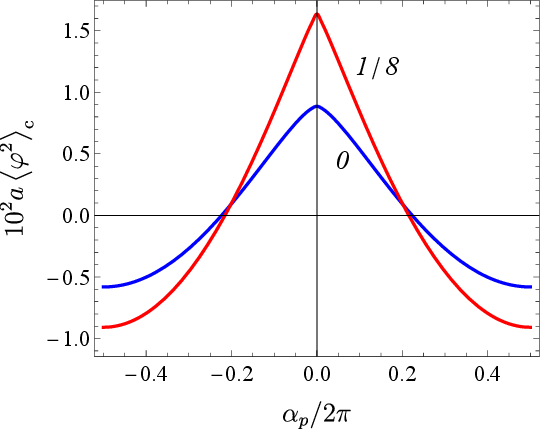,width=7.5cm,height=6cm} & \quad{}\epsfig{figure=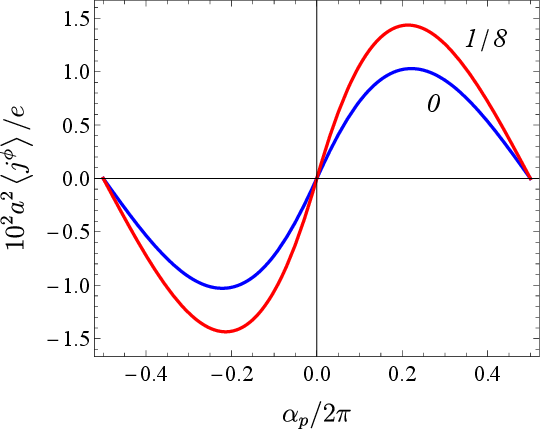,width=7.5cm,height=6cm}\tabularnewline
\end{tabular}
\par\end{centering}
\caption{The VEV of field squared (left panel) and the azimuthal current density
(right panel) as a function of the phase in the quasiperiodicity condition
for conformally ($\xi=1/8$) and minimally ($\xi=0$) coupled fields
with $L/r=0.5$ and $ma=0.5$. The numbers near the curves are the values of $\protect\xi .$}
\label{fig4}
\end{figure}

It would be useful to compare numerically the compact counterpart
of the VEV of field squared (\ref{FldSq2}) with the corresponding
azimuthal current density (\ref{jphicc3}). The VEVs $\left\langle \varphi^{2}\right\rangle _{\mathrm{c}}$
and $\langle j^{\phi}\rangle$ versus the radial coordinate in the
form of the ratio $L/r$ are shown in the left and right panels of
Fig. \ref{fig2}. The graphs are plotted for conformally and minimally
coupled massless fields assuming that the phase in the quasiperiodicity
condition is equal to $\pi/2$. The numbers near curves represent
the values of the curvature coupling parameter $\xi$. Both VEVs diverge
and have opposite signs for large values of $r/L$ (the VEV of field
squared diverges as $r/L$, whereas $\langle j^{\phi}\rangle$ is
increasing faster, as $(r/L)^{2}$).
In Fig. \ref{fig3} the same quantities are presented versus field
mass in the units of $a^{-1}$ for the values of parameters $\alpha_{p}=2\pi/5$
and $L/r=0.5$. Both VEVs are vanishing for large masses, however
$\left\langle \varphi^{2}\right\rangle _{\mathrm{c}}$ has a maximum
value for nonzero field mass.
The same VEVs are depicted in Fig. \ref{fig4}, as functions of the
parameter $\alpha_{p}/2\pi$ for conformally and minimally coupled
fields assuming that $L/r=0.5$ and $ma=0.5$. As mentioned above,
the VEV of field squared is an even function of the phase, whereas
the VEV of the current density is an odd function.

\section{VEV of the energy-momentum tensor}

\label{sec:emt}

As another important characteristic of the vacuum state we consider
the VEV of energy-momentum tensor. The latter can be obtained with
the help of the formula \cite{Birr82}
\begin{equation}
\left\langle T_{ik}\right\rangle =\frac{1}{4}\underset{x^{\prime}\rightarrow x}{\lim}\left(\partial_{i^{\prime}}\partial_{k}+\partial_{k^{\prime}}\partial_{i}\right)G\left(x,x^{\prime}\right)+\left(\xi-\frac{1}{4}\right)g_{ik}\nabla_{p}\nabla^{p}\left\langle \varphi^{2}\right\rangle -\xi\nabla_{i}\nabla_{k}\left\langle \varphi^{2}\right\rangle -\xi R_{ik}\left\langle \varphi^{2}\right\rangle ,\label{Tik}
\end{equation}
where $g_{ik}$ is the metric tensor and $\nabla_{i}$ is the covariant
derivative. The straightforward calculation with the use of the expressions
(\ref{Gcc}) and (\ref{FldSq}) gives the following expressions for
the nonzero components (no summation over $i$):
\begin{align}
\left\langle T_{i}^{i}\right\rangle  & =\frac{1}{2\pi a^{3}}\int_{0}^{\infty}d\nu\,\frac{\nu}{\sqrt{\nu^{2}+\nu_{m}^{2}}}\tanh\left(\pi\nu\right)F_{i}^{(0)}\left(\nu\right)\nonumber \\
 & +\frac{1}{2\pi a^{3}}\sum_{l=1}^{\infty}\cos\left(l\alpha_{p}\right)\int_{0}^{\infty}d\nu\,\frac{\nu}{\sqrt{\nu^{2}+\nu_{m}^{2}}}\tanh\left(\pi\nu\right)\hat{F}_{i}P_{i\nu-1/2}\left(x\right),\label{Tii}
\end{align}
where
\begin{equation}
F_{0}^{(0)}\left(\nu\right)=\frac{1}{2}\left(\nu^{2}+\nu_{m}^{2}\right),\:F_{1}^{(0)}\left(\nu\right)=F_{2}^{(0)}\left(\nu\right)=\frac{1}{4}\left(-\nu^{2}+2\xi-\frac{1}{4}\right),\label{Fnu}
\end{equation}
 and the operators $\hat{F}_{i}$, $i=0,1,2$, are defined as
\begin{align}
\hat{F}_{0} & =-2\left(\xi-\frac{1}{4}\right)\left(x-1\right)\left[2\left(x-1\right)\partial_{x}^{2}+3\partial_{x}\right]+\nu^{2}+\nu_{m}^{2},\label{F0}\\
\hat{F}_{1} & =-\frac{1}{2}\left[\left(4\xi-1\right)x-4\xi-1\right]\partial_{x}+\xi,\label{F1}\\
\hat{F}_{2} & =-\frac{1}{2}\left\{ 2\left(x-1\right)\left[\left(4\xi-1\right)x-4\xi-1\right]\partial_{x}^{2}+\left[\left(8\xi-3\right)x-8\xi+1\right]\partial_{x}-2\xi\right\} .\label{F2}
\end{align}
Similar to the case of the field squared, the vacuum energy-momentum
tensor is an even function of $\alpha_{p}$. The VEV of energy-momentum
tensor (\ref{Tii}) obeys the trace relation
\begin{equation}
\left\langle T_{k}^{k}\right\rangle =\left[2\left(\xi-1/8\right)\nabla_{k}\nabla^{k}+m^{2}\right]\left\langle \varphi^{2}\right\rangle .\label{trrel}
\end{equation}
The covariant conservation equation $\nabla_{k}\left\langle T_{i}^{k}\right\rangle =0$
with $i=1$ gives the relation
\begin{equation}
\left\langle T_{2}^{2}\right\rangle =\left(1-r\frac{\partial}{\partial r}\right)\left\langle T_{1}^{1}\right\rangle ,\label{cceqn}
\end{equation}
which is satisfied by the expression (\ref{Tii}). The remaining components
of the conservation equation are reduced to a trivial identity.

An alternative form for the VEVs can be obtained by means of the expressions
(\ref{Gccn4}) and (\ref{FldSq1}):
\begin{align}
\left\langle T_{i}^{i}\right\rangle  & =\frac{1}{\pi^{2}a^{3}}\int_{\nu_{m}}^{\infty}d\nu\,\frac{\nu}{\sqrt{\nu^{2}-\nu_{m}^{2}}}R_{i}^{(0)}\left(\nu\right)\underset{x\rightarrow1}{\lim}Q_{\nu-1/2}\left(x\right)+\left\langle T_{i}^{i}\right\rangle _{\mathrm{c}},\label{Tii-1}
\end{align}
with
\[
R_{0}^{(0)}\left(\nu\right)=F_{0}^{(0)}\left(\nu\right)-\nu^{2},\:R_{1}^{(0)}\left(\nu\right)=R_{2}^{(0)}\left(\nu\right)=F_{1}^{(0)}\left(\nu\right)+\frac{\nu^{2}}{2},
\]
and with the compactified part $\left\langle T_{i}^{i}\right\rangle _{\mathrm{c}}$
having the form
\begin{equation}
\left\langle T_{i}^{i}\right\rangle _{\mathrm{c}}=\frac{1}{\pi^{2}a^{3}}\sum_{l=1}^{\infty}\cos\left(l\alpha_{p}\right)\int_{\nu_{m}}^{\infty}d\nu\,\frac{\nu}{\sqrt{\nu^{2}-\nu_{m}^{2}}}\hat{R}_{i}Q_{\nu-1/2}\left(x\right).\label{Tii-2}
\end{equation}
Here the operators$\hat{R}_{i}$, $i=0,1,2$, are defined as
\begin{equation}
\hat{R}_{i}=\hat{F}_{i}-2\nu^{2}\delta_{0i},\label{Ri}
\end{equation}
with $\hat{F}_{i}$ given by (\ref{F0})-(\ref{F2}) and $\delta_{0i}$
being the Kronecker delta. Note that relation (\ref{cceqn}) is verified
for the final expressions (\ref{Tii-1}), by taking into account that%
\begin{equation}
\frac{\partial }{\partial r}\underset{x\rightarrow 1}{\lim }Q_{\nu
-1/2}\left( x\right) =0,  \label{ddrQ}
\end{equation}%
and%
\begin{equation}
1-r\frac{\partial }{\partial r}=1+2\left( x-1\right) \frac{\partial }{%
\partial x}.  \label{ddrddx}
\end{equation}

As an important special case we consider the conformally coupled massless
field. In this case one has $\nu_{m}=0$, $\xi=1/8$ and the VEVs
are given as
\begin{align}
\left\langle T_{i}^{i}\right\rangle  & =\frac{1}{2\pi a^{3}}\int_{0}^{\infty}d\nu\,\tanh\left(\pi\nu\right)F_{i,\mathrm{c}}^{(0)}\left(\nu\right)+\frac{1}{2\pi a^{3}}\sum_{l=1}^{\infty}\cos\left(l\alpha_{p}\right)\int_{0}^{\infty}d\nu\,\tanh\left(\pi\nu\right)\hat{F}_{i,\mathrm{c}}P_{i\nu-1/2}\left(x\right)\nonumber \\
 & =\frac{1}{\pi^{2}a^{3}}\int_{0}^{\infty}d\nu\,R_{i,\mathrm{c}}^{(0)}\left(\nu\right)\underset{x\rightarrow1}{\lim}Q_{\nu-1/2}\left(x\right)+\frac{1}{\pi^{2}a^{3}}\sum_{l=1}^{\infty}\cos\left(l\alpha_{p}\right)\int_{0}^{\infty}d\nu\,\hat{R}_{i,\mathrm{c}}Q_{\nu-1/2}\left(x\right),\label{Tii-3}
\end{align}
with 
\begin{equation}
F_{0,\mathrm{c}}^{(0)}\left(\nu\right)=-2F_{1,\mathrm{c}}^{(0)}\left(\nu\right)=-2F_{2,\mathrm{c}}^{(0)}\left(\nu\right)=-R_{0,\mathrm{c}}^{(0)}\left(\nu\right)=2R_{1,\mathrm{c}}^{(0)}\left(\nu\right)=2R_{2,\mathrm{c}}^{(0)}\left(\nu\right)=\frac{\nu^{2}}{2},\label{Fnu-1}
\end{equation}
and
\begin{align*}
\hat{F}_{0,\mathrm{c}} & =\hat{R}_{0,\mathrm{c}}+2\nu^{2}=\frac{1}{4}\left(x-1\right)\left[2\left(x-1\right)\partial_{x}^{2}+3\partial_{x}\right]+\nu^{2},\\
\hat{F}_{1,\mathrm{c}} & =\hat{R}_{1,\mathrm{c}}=\frac{1}{8}\left[2\left(x+3\right)\partial_{x}+1\right],\\
\hat{F}_{2,\mathrm{c}} & =\hat{R}_{2,\mathrm{c}}=\frac{1}{8}\left[4\left(x-1\right)\left(x+3\right)\partial_{x}^{2}+8x\partial_{x}+1\right].
\end{align*}
Note that the geometry given by (\ref{ds2}) is conformally related
to the 2D Rindler spacetime described by the line element $ds_{\mathrm{R}}^{2}=r^{2}d\tau^{2}-dr^{2}-dy^{2}$
with dimensionless time coordinate $\tau=t/a$ and compact coordinate
$y=L\phi$ (for the VEVs in $(D+1)$-dimensional Rindler spacetime
with general number of compact dimensions see \cite{Kota22}), i.e.
$ds^{2}=\left(a/r\right)^{2}ds_{\mathrm{R}}^{2}$. Thus, the diagonal
components of the topological contributions in the VEV of energy-momentum
tensor in the locally Rindler spacetime, $\left\langle T_{k}^{k}\right\rangle _{\mathrm{R},\mathrm{c}}$,
are related to the components $\left\langle T_{k}^{k}\right\rangle _{\mathrm{c}}$
by the formulae (no summation over $k$)
\begin{equation}
\left\langle T_{k}^{k}\right\rangle _{\mathrm{R},\mathrm{c}}=\left(a/r\right)^{3}\left\langle T_{k}^{k}\right\rangle _{\mathrm{c}},\:k=0,1,2.\label{EMTR}
\end{equation}

Now, we discuss the asymptotic behavior of the compact counterparts
in the VEV of energy-momentum tensor for the limiting cases of the
ratio $r/L$. For $r/L\ll1$ one has the approximations
\begin{align}
\left\langle T_{0}^{0}\right\rangle _{\mathrm{c}} & \approx\left(1-4\xi\right)\nu_{m}\left(\nu_{m}+1/2\right)T\left(2\pi L/r\right),\label{SmrT00}\\
\left\langle T_{1}^{1}\right\rangle _{\mathrm{c}} & \approx\left[\left(2\xi-1/2\right)\left(\nu_{m}+1/2\right)+\xi\right]T\left(2\pi L/r\right),\label{SmrT11}\\
\left\langle T_{2}^{2}\right\rangle _{\mathrm{c}} & \approx\left\{ \xi-\left[4\xi\left(\nu_{m}+1/2\right)-\nu_{m}\right]\left(\nu_{m}+1/2\right)\right\} T\left(2\pi L/r\right),\label{SmrT22}
\end{align}
with the function $T\left(z\right)$ defined as
\begin{equation}
T\left(z\right)=\frac{\sqrt{\nu_{m}}\Gamma\left(\nu_{m}+1/2\right)}{2\pi a^{3}\Gamma\left(\nu_{m}+1\right)}\frac{z^{-2\nu_{m}-1}}{\sqrt{\ln z}}\sum_{l=1}^{\infty}\frac{\cos\left(l\alpha_{p}\right)}{l^{2\nu_{m}+1}}.\label{Tz}
\end{equation}
In the case of a conformally coupled massless field the corresponding
leading terms have the form
\begin{equation}
\left\langle T_{0}^{0}\right\rangle _{\mathrm{c}}\approx-\left\langle T_{1}^{1}\right\rangle _{\mathrm{c}}\approx\ln\left(2\pi L/r\right)\left\langle T_{2}^{2}\right\rangle _{\mathrm{c}}\approx\frac{r/L}{32\pi^{2}a^{3}\left[\ln\left(2\pi L/r\right)\right]^{2}}\sum_{l=1}^{\infty}\frac{1}{l}\cos\left(l\alpha_{p}\right).\label{SmrTiic}
\end{equation}
The approximations for the opposite limit $r/L\gg1$ are obtained
similar to (\ref{lgrfldsq}) and are given as
\begin{equation}
\left\langle T_{0}^{0}\right\rangle _{\mathrm{c}}\approx\left\langle T_{1}^{1}\right\rangle _{\mathrm{c}}\approx-\frac{1}{2}\left\langle T_{2}^{2}\right\rangle _{\mathrm{c}}\approx-\frac{\left(r/L\right)^{3}}{16\pi^{4}a^{3}}\sum_{l=1}^{\infty}\frac{\cos\left(l\alpha_{p}\right)}{l^{3}},\label{lgrTii}
\end{equation}
which coincide with the VEVs for a cylindrical tube with constant
radius $aL/r$. Similar to the field squared, the dominant contribution in the
topological VEV of the energy--momentum tensor comes from high-momentum modes
in the compactified dimension and the curvature effects are relatively weak.
The leading term in the asymptotic expansion behaves as $\left( r/L\right)
^{3}$, which indicates an unbounded growth of vacuum polarization due to the
topology. In the semiclassical approach that we have adopted, the
renormalized vacuum energy--momentum tensor stands on the right hand side of
the Einstein field equations. Thus, the unbounded growth of VEVs at small
proper radius of the compactified dimension can act as a source of
significant curvature, implying that the back-reaction effects on the
geometry could become significant.

\begin{figure}[tbph]
\begin{centering}
\begin{tabular}{cc}
\epsfig{figure=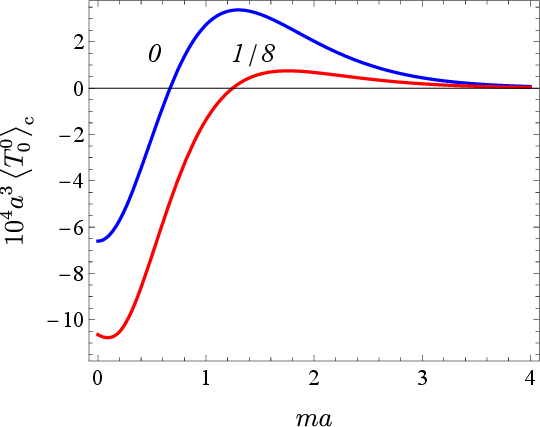,width=7.5cm,height=6cm} & \quad{}\epsfig{figure=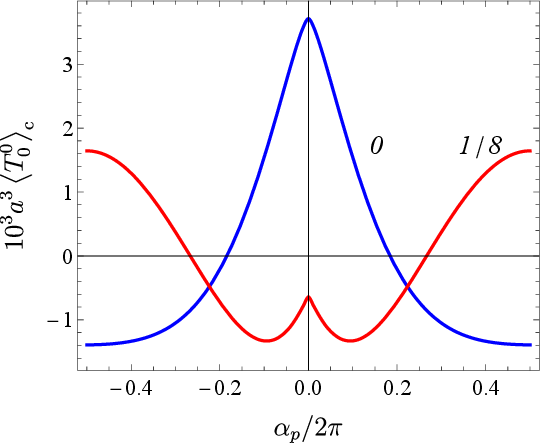,width=7.5cm,height=6cm}\tabularnewline
\end{tabular}
\par\end{centering}
\caption{The energy density as a function of the mass (left panel) and of
the phase in the quasiperiodicity condition (right panel) for conformally ($%
\protect\xi =1/8$) and minimally ($\protect\xi =0$) coupled fields and for $%
L/r=0.5$.
For the left panel we have taken $\alpha_{p}=2\pi/5$ and for the
right panel $ma=0.5$. The numbers near the curves are the values of $\protect\xi .$}
\label{fig5}
\end{figure}

\begin{figure}[tbph]
\begin{centering}
\begin{tabular}{cc}
\epsfig{figure=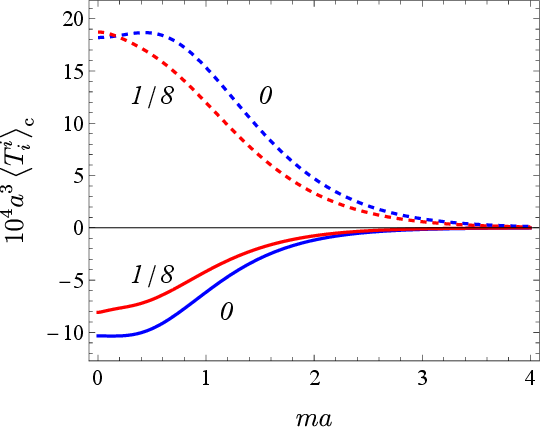,width=7.5cm,height=6cm} & \quad{}\epsfig{figure=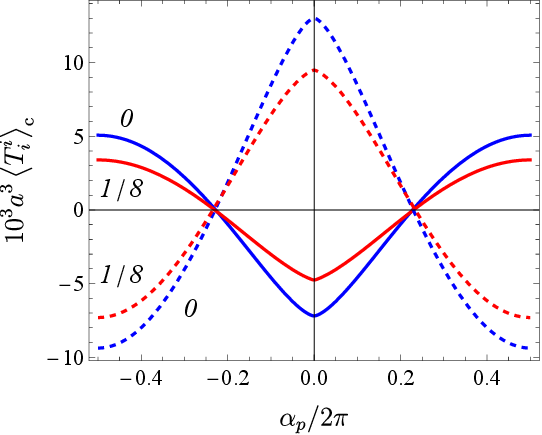,width=7.5cm,height=6cm}\tabularnewline
\end{tabular}
\par\end{centering}
\caption{Same as in Fig. \ref{fig5} for $\left\langle T_{1}^{1}\right\rangle _{\mathrm{c}}$
(full curves) and $\left\langle T_{2}^{2}\right\rangle _{\mathrm{c}}$
(dashed curves).}
\label{fig6}
\end{figure}

\begin{figure}[tbph]
\begin{centering}
\begin{tabular}{cc}
\epsfig{figure=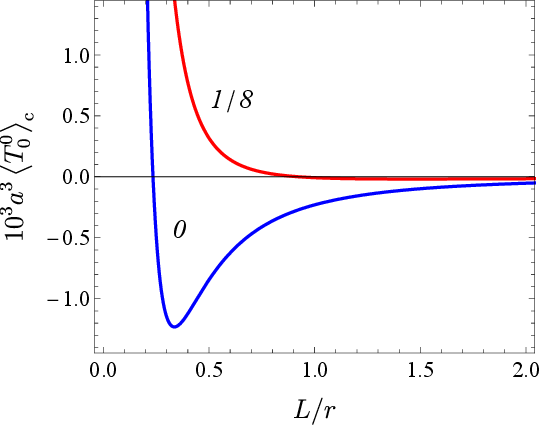,width=8cm,height=7cm} & \quad{}\epsfig{figure=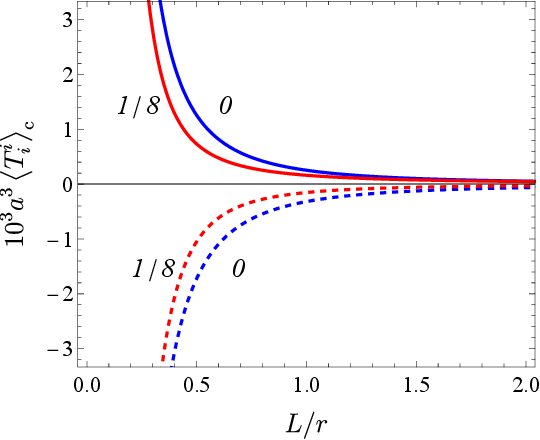,width=8cm,height=7cm}\tabularnewline
\end{tabular}
\par\end{centering}
\caption{The vacuum energy density (left panel) and vacuum stresses (right
panel) versus $L/r$ for conformally ($\xi=1/8$) and minimally ($\xi=0$)
coupled massless fields and for fixed $\alpha_{p}=\pi/2$. Full and
dashed curves on the right panel correspond to $\left\langle T_{1}^{1}\right\rangle _{\mathrm{c}}$
and $\left\langle T_{2}^{2}\right\rangle _{\mathrm{c}}$, respectively. The numbers near the curves are the values of $\protect\xi .$}
\label{fig7}
\end{figure}

Here we discuss some numerical examples presented for the compact
counterpart in the first diagonal component of the VEV of the energy-momentum
tensor corresponding to the energy density. The left panel of Fig.
\ref{fig5} presents the dependence of the energy density on the mass
of the field for the parameters $L/r=0.5$ and $\alpha_{p}=2\pi/5$.
The right panel displays the energy density versus $\alpha_{p}/2\pi$
for the same value of $L/r$ and for $ma=0.5$. On both panels the
numbers near the curves are the corresponding values of the curvature
coupling parameter. The corresponding dependences for vacuum stresses
$\left\langle T_{i}^{i}\right\rangle _{\mathrm{c}}$, $i=1,2$ (no
summation over $i$) are depicted in Fig. \ref{fig6}. Here, the full
and dashed curves correspond to $\left\langle T_{1}^{1}\right\rangle _{\mathrm{c}}$
and $\left\langle T_{2}^{2}\right\rangle _{\mathrm{c}}$, respectively.
In Fig. \ref{fig7}, the left panel shows the dependence of the energy
density on the ratio $L/r$ for minimally and conformally coupled
massless fields. The graphs are plotted for $\alpha_{p}=\pi/2$. As
was expected, the energy density tends to zero in the limit $L/r\rightarrow\infty$.
The corresponding dependences for vacuum stresses are presented on
the right panel. Again, the full (dashed) curves represent the radial
(azimuthal) stress.

\section{Conclusion}

\label{sec:conc}

We have investigated the combined effects of spatial curvature and
topology on the properties of the vacuum state for a charged scalar
field localized on the (2+1)-dimensional Beltrami pseudosphere. The
corresponding geometry is described by the line element (\ref{ds2}).
It is assumed that the field obeys the quasiperiodicity condition
(\ref{pc}) where the phase is constant. With the help of the expressions
obtained in \cite{Saha24} for the two-point Hadamard function, two
equivalent representations are obtained for the important physical
characteristics of the vacuum state. As such characteristics we have
considered the VEVs of the field squared and of the energy-momentum
tensor. The corresponding expressions are given by (\ref{FldSq})
and (\ref{Tii}) for the first representation, and (\ref{FldSq1})
and (\ref{Tii-1}) for the second one. The VEVs are decomposed into
compactified and uncompactified parts. The latter parts are divergent,
whereas the compact counterparts are finite. Thus, the renormalization
of the VEVs is reduced to that for the uncompactified parts only.
As an important special case we have discussed the conformally coupled
massless scalar field. The geometry for the Beltrami pseudosphere
is conformally related to the (2+1)-dimensional Rindler spacetime
and the corresponding VEVs of the energy-momentum tensor in these
two spacetimes are related by the formula (\ref{EMTR}) for a conformally
coupled massless field.

In addition, the topological contributions have been analysed asymptotically
for the limiting values of the ratio $r/L$. For a given $r$ the
dimensionless ratio $r/L$ corresponds to the inverse of the proper
radius of the compactified dimension measured in units of the curvature
radius. In the case of $r/L\ll1$ the VEVs are approximated as (\ref{smrfs})
and (\ref{SmrT00})-(\ref{SmrT22}). In this limit, the decay of the
compact counterpart in the energy density as a function of $r/L$
follows a power-law, as $\left(r/L\right)^{2\nu_{m}+1}$. The formulae
(\ref{SmrT00})-(\ref{SmrT22}) are not applicable for the conformally
coupled massless case and the expressions (\ref{SmrTiic}) are provided
for that case. Contrary to the VEV of the field squared and the energy
density, the absolute values of the radial and azimuthal stresses
are increasing as the ratio $r/L$ tends to zero, therefore the effect
of nontrivial topology is strong for the stresses at small values
of the radial coordinate in the conformally coupled massless case.
The nontrivial topology is essential also in the opposite asymptotic
limit, $r/L\gg1$. The corresponding leading terms are given as (\ref{lgrfldsq})
and (\ref{lgrTii}). The approximate expressions are independent of
the mass $m$ and of the curvature coupling parameter $\xi$. The
magnitudes of the VEVs are increasing by a power-law as the ratio
$r/L$ takes large values. In the conformally coupled massless
case, the growth of the radial and azimuthal stresses is a manifestation of
the small-compactification amplification that is universal to topological
Casimir systems (even in the locally Minkowski bulk). These divergences stem
from vacuum fluctuations with high frequencies, and they are analogous to
ultraviolet divergences in quantum field theories on the Minkowski bulk. In
the present geometry, the proper radius of the compactified dimension
decreases in the regime  $r/L\gg 1$, leading to the observed amplification.
As with standard Casimir problems, the resulting vacuum expectation values
(VEVs) may violate classical energy conditions, a generic feature of quantum
vacuum polarization. The back-reaction effects become relevant when the VEV
of the energy--momentum tensor, multiplied by Newton's gravitational
constant, approaches the background curvature scale.

\section*{Acknowledgments}

The research was supported by the Higher Education and Science Committee
of MESCS RA (Research projects No. 21AG-1C047 and No. 24FP-3B021).

\end{document}